\documentclass[twocolumn]{aastex63}
\usepackage{graphicx}
\usepackage{amsmath}
\usepackage{amssymb}
\usepackage{xspace,xcolor}
\usepackage{natbib}
\usepackage{outlines}
\usepackage{savesym}
\usepackage{booktabs}
\usepackage{longtable}
\usepackage{tikz}
\usepackage{graphicx}
\usepackage{hyperref}
\savesymbol{tablenum}
\usepackage{siunitx}
\usepackage{comment}
\usepackage{float}
\usepackage{braket}

\newcommand{\be}{\begin{eqnarray}}
\newcommand{\ee}{\end{eqnarray}}
\newcommand{\lp}{\left(}
\newcommand{\rp}{\right)}
\newcommand{\lb}{\left[}
\newcommand{\rb}{\right]}

\graphicspath{{./}{figures/}}

\newcommand{\OCIS}{The Observatories of the Carnegie Institution for Science, Pasadena, CA 91101, USA}
\newcommand{\JHU}{William H. Miller III Department of Physics \& Astronomy, Johns Hopkins University, 3400 N Charles St, Baltimore, MD 21218, USA}
\newcommand{\ARCO}{Astrophysics Research Center of the Open University (ARCO), Department of Natural Sciences, Ra’anana 4353701, Israel}

\begin{document}
\shorttitle{}
\shortauthors{Piro, Zenati \& Wong}

\title{Thermal Evolution of the Central Star in Pa~30}

\correspondingauthor{Anthony L.\ Piro}
\email{piro@carnegiescience.edu}

\author[0000-0001-6806-0673]{Anthony L.\ Piro}
\affiliation{\OCIS}

\author[0000-0002-0632-8897]{Yossef\ Zenati}
\affiliation{\ARCO}
\affiliation{\JHU}

\author[0000-0001-9195-7390]{Tin Long Sunny\ Wong}
\affiliation{\OCIS}

\begin{abstract}
Pa~30 has been identified as the nebular remnant of the historical SN~1181. It is host to a hot \mbox{($\approx200,000\,{\rm K}$)} central star (WD~J005311) with a fast wind ($\approx16,000\,{\rm km\,s^{-1}}$) radiating at roughly the Eddington luminosity for a solar mass ($\approx1.5\times10^{38}\,{\rm erg\,s^{-1}}$). We explore the thermal evolution of this star to understand how it progressed toward the state it is observed as today as well as to constrain its underlying physical properties. We develop a semi-analytic two-component model, which approximates the central star as a hot radiating envelope contracting and cooling above a relatively cool core. Comparing this model with the observed luminosity and radius requires a core mass $M_c\approx1.15-1.4\,M_\odot$ with a core radius $R_c\approx(6-8)\times10^8\,M_\odot$, and a hot envelope mass $\Delta M\approx0.02-0.04\,M_\odot$. The small envelope mass is the best constrained of these parameters due to the need to reach the observed radius of $\approx0.15\,R_\odot$ in a timescale of $\approx845\,{\rm yrs}$. These results favor a picture where SN~1181 involved the merger of O/Ne and C/O white dwarfs, and where the majority of the latter was ejected in the explosion. We also explore which models ignite carbon burning at the base of the hot envelope, demonstrating that this is possible but not necessarily required to explain the current thermal state of the central star.
\end{abstract}


\section{Introduction}
\label{sec:intro}

IRAS~00500+6713 (Pa~30) was identified in the WISE mid-IR archive by Dana Patchick \citep{CutriR+2012,KronbergerM+2016} and cataloged in HASH \citep{ParkerQ2017}. Follow-up observations revealed an extremely hot and luminous central source (WD~J005311) devoid of H and He, with emission lines in C and O that imply a wind velocity of $\approx16,000\,{\rm km\,s^{-1}}$ and a mass loss rate of $\sim 10^{-6}\,M_\odot\,\mathrm{yr}^{-1}$ \citep{GvaramadzeV+19_Nat}. Subsequent work connected Pa~30 to the historical SN~1181, implying a current age of $\approx 845\,{\rm yrs}$ \citep{Ritter2021,Lykou23,FasenR+23,Schaefer23}. Mapping of filaments reveals ballistic ejecta at $\sim 10^3\,\mathrm{km\,s^{-1}}$, a large inner cavity, and a sharply bounded, filamentary shell, largely consistent with the 1181 explosion date \citep{CunninghamT+2024}.

Due to the relatively dim maximum brightness, together with the low expansion velocities, it has been argued that Pa~30 was the result of a subluminous class of Type~Ia supernovae, called  SNe~Iax \citep{Li2003,Foley2013,Jha2017}. This is also consistent with X-ray studies, which reveal large neon, magnesium, silicon, and sulfur enrichment of the central star and the nebula \citep{OskinovaL+20} and requires a low explosion energy \citep{KoT+24,Shao2025}. A further connection is that SNe~Iax do not transition to a nebular phase like normal SN~Ia \citep[e.g.,][]{Foley2016,Kawabata2018}, likely because these explosions leave some sort of bound central star with a wind \citep[which could be driven by delayed radioactive decay for the first few years to decades,][]{Shen2017}.

The nebula composition and near-Eddington luminosity favors thermonuclear burning in a roughly solar-mass system, but the outflow speeds of $\approx16,000\,{\rm km\,s^{-1}}$ far exceed the escape velocity of a white dwarf (WD). This disfavors a purely radiatively-driven wind, and points to a magnetocentrifugal flinging \citep{Kashiyama19,Zhong24}. A natural way to produce such a fast spin $\sim0.1\,{\rm s}^{-1}$ and strong magnetic field $\sim10^8\,{\rm G}$ is via a double degenerate merger \citep[e.g.,][]{Garcia12,Schwab12,KilicM+21}. This fits with growing evidence that at least some SNe~Iax occur from mergers \citep[e.g.,][]{Karambelkar2021}, while others are associated with WDs with stellar companions \citep{McCully2014,Foley2014}. In still other cases, only limits could be placed on a companion \citep{Foley2015,Zimmerman2026}, so low-mass main-sequence or He-star stars companion, as well as double degenerate systems, are all possible.

From the above collection of properties for Pa~30, a general picture is developing. A double degenerate merger occurred, leading to a non-terminal explosion that ejected many tenths of a solar mass in a low energy outburst. The burning likely proceeded as a deflagration that initiated somewhere near the merged product of the two WDs. The current configuration is thus a hot layer consisting of some combination of the burned and unburned material from the less massive WD as well as fallback material that was not completely unbound in the low energy explosion \citep[e.g.,][which find that $\sim0.01-0.2\,M_\odot$ of burned material may be retained in a deflagration]{Jordan2012,Kromer2013,Fink2014,Long2014}. This sits on top of a relatively cooler core of the more massive WD, although this may have been heated somewhat depending on the details of where and how the deflagration took place. Such configurations of mass and energy have been studied a number of times \citep[e.g.,][]{Schwab12,Schwab2021,WuC+22,Yao2023,Ko2026}. Here we approach this problem with a semi-analytic framework \citep[similar to the so-called Eddington model but applied to a layer, e.g.,][]{Eddington1926,Hansen1994}. This allows us to explore a wide parameter space of models and better understand what underlying physical characteristics of the central star are constrained by the observed properties.

In Section~\ref{sec:model}, we present our semi-analytic model for the Pa~30 central star along with an exploration varying various properties of the envelope and core lead to different observational consequences. In Section~\ref{sec:c12}, we discuss whether carbon burning is expected to be ignited in the hot surface layers and what impact it may have on the evolution. In Section~\ref{sec:pa30}, we explore what properties of the Pa~30 central star can be best constrained through this modeling, and in Section~\ref{sec:conclusion}, we summarize our results as well as discuss extensions of this work for the future. 

\section{Hot Envelope Model}
\label{sec:model}

\begin{figure}
\includegraphics[width=0.45\textwidth,trim=1.0cm 1.0cm 0.5cm 0.5cm]{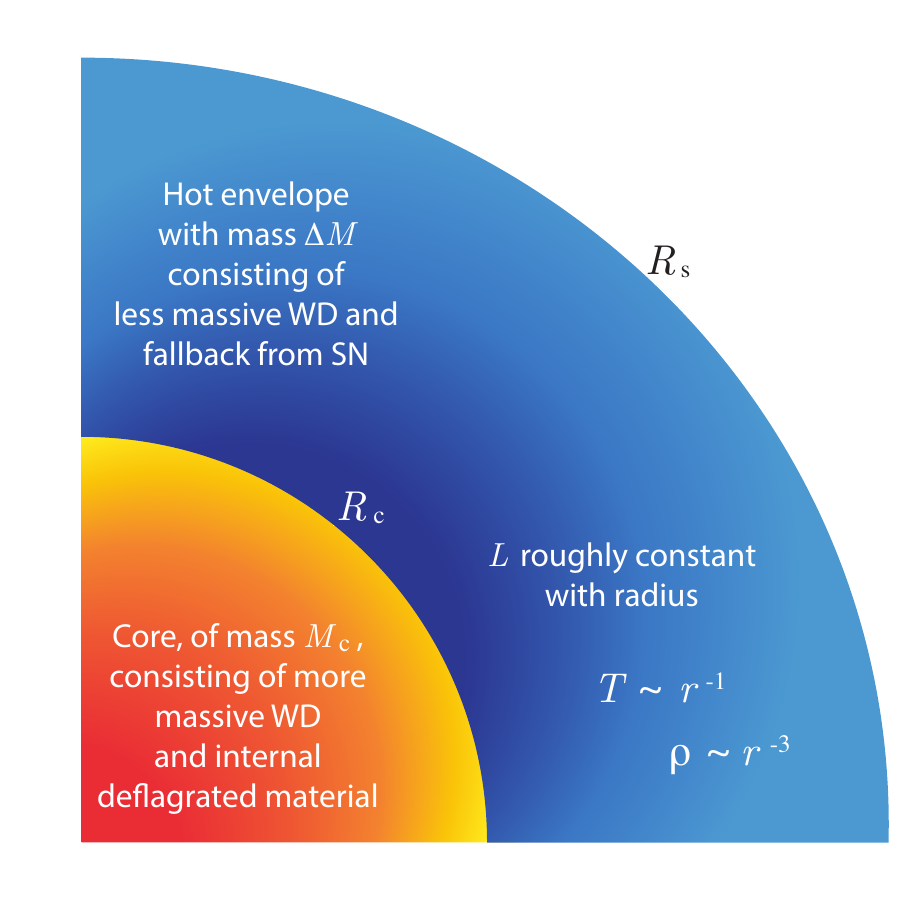}
\caption{Diagram of a quadrant of the central star in Pa~30. The blue region represents the hot envelope with mass $\Delta M$ and the orange/red region represents the cooler core with mass $M_c$. As we derive in Section~\ref{sec:hot envelope}, for a constant luminosity $L$ though the envelope and energy carried by radiative diffusion, this layer has a $T\propto r^{-1}$ and $\rho\propto r^{-3}$ profile. As $L$ goes down with time, the surface radius $R_s$ contracts as the layer heats.}
\label{fig:diagram}
\end{figure}

Our general picture of the central star of Pa~30 is summarized in Figure~\ref{fig:diagram} and consists of a hot envelope with mass $\Delta M$ and radius $R_s$, which sits above a core with mass $M_c$ and radius $R_c$. The core represents the more massive WD from the merger that produced SN~1181. The radius $R_c$ can in general be larger than simply a zero-temperature WD of the same mass due to the heating from the merger event plus the deflagration that produced the SN~Iax. The hot envelope is a combination of the material from the less massive WD that was completely disrupted in the event (minus the fraction of this WD that was ejected in the explosion) plus any fallback that occurred shortly after the explosion (since the explosion was so low energy). We next describe a toy model that can replicate the main features of the expected thermal evolution of such a configuration, and explore how the evolution depends on the main properties of the system, namely $M_c$, $R_c$, and $\Delta M$.

\subsection{Hot Envelope Model Framework}
\label{sec:hot envelope}

Flux limited diffusion through the envelope requires
\be
	L = -\frac{16\pi r^2acT^3}{3\kappa\rho} \frac{dT}{dr},
	\label{eq:radiative diffusion}
\ee
where $a$ is the radiation constant and $\kappa$ is the opacity. Hydrostatic equilibrium is given by $dP/dr = -\rho GM_c/r^2$, where we approximate $\Delta M\ll M_c$ (this is found to be consistent with what we constrain for Pa~30). Combining this with Equation~(\ref{eq:radiative diffusion}) results in
\be
	L = \frac{16 \pi GM_cacT^3}{3\kappa}\frac{dT}{dP}.
	\label{eq:radiative diffusion2}
\ee
The luminosity will generally be near the Eddington luminosity
\be
	L_{\rm Edd} = \frac{4\pi GM_c c}{\kappa},
\ee
so it is helpful to parameterize the envelope solutions by the Eddington ratio $\chi\equiv L/L_{\rm Edd}$\footnote{Note that one can also interpret $\chi$ as the ratio of radiation pressure to total pressure, so that $\chi=aT^4/(3P)$ and \mbox{$1-\chi = \rho k_{\rm B}T/(\mu m_p P)$.}}. Integrating Equation~(\ref{eq:radiative diffusion2}) with $L$ set constant results in a temperature profile
\be
	T(P) = \chi^{1/4}\lp\frac{3P}{a} \rp^{1/4}.
	\label{eq:temperature}
\ee
We assume that $\kappa$ is constant because it is roughly given by electron scattering in the hot envelope.

For the equation of state, we consider contributions from both ideal gas and radiation, so that
\be
	P = \frac{\rho k_{\rm B}T}{\mu m_p}
    + \frac{aT^4}{3},
    \label{eq:pressure1}
\ee
where $k_{\rm B}$ is the Boltzmann constant, $\mu$ is the mean molecular weight, and $m_p$ is the proton mass. Combining this with Equation~(\ref{eq:temperature}), results in pressure as a function of density of
\be
	P(\rho) = \lp \frac{k_{\rm B}}{\mu m_p} \rp^{4/3} \lp  \frac{3}{a} \rp^{1/3} \frac{\chi^{1/3}}{(1-\chi)^{4/3}} \rho^{4/3}.
    \label{eq:pressure2}
\ee
This demonstrates that the envelope is represented by an $n=3$ polytrope. This means we can simply use $P=K\rho^{4/3}$, where
\be
	K\equiv \lp \frac{k_{\rm B}}{\mu m_p} \rp^{4/3} \lp  \frac{3}{a} \rp^{1/3} \frac{\chi^{1/3}}{(1-\chi)^{4/3}},
\ee
is a factor (the entropy) that increases as the envelope gets near the Eddington limit.

We can also solve for the thermodynamic properties as a function of radius by using the radiative diffusion equation. Substituting
\be
	\rho(T) = \frac{\mu m_p}{k_{\rm B}}\frac{aT^3}{3}\frac{1-\chi}{\chi}.
\ee
into Equation~(\ref{eq:radiative diffusion}) results in
\be
	\frac{dT}{dr} = - \frac{\mu m_p}{k_{\rm B}} \frac{GM_c}{4r^2}(1-\chi),
\ee
and integrating this results in
\be
	T(r) = \frac{\mu m_p}{k_{\rm B}} \frac{GM_c}{4} (1-\chi)
    \lp \frac{1}{r} - \frac{1}{R_s} \rp.
\ee
Similarly,
\be
	\rho(r) = \frac{a}{3} \lp\frac{\mu m_p}{k_{\rm B}}\rp^4 \lp \frac{GM_c}{4} \rp^3 \frac{(1-\chi)^4}{\chi}
    \lp \frac{1}{r} - \frac{1}{R_s} \rp^3.
    \nonumber
    \\
    \label{eq:rho}
\ee
for the density profile within the envelope.

The envelope radius $R_s$ is found by requiring the envelope to have the correct mass
\be
	\Delta M = \int_{R_c}^{R_s} 4\pi r^2\rho dr.
\ee
Substituting $\rho(r)$ from Equation~(\ref{eq:rho}) and integrating this expression results in the relationship
\be
    \Delta M = &&4\pi B(\chi)
    \left\{
        \ln\lp\frac{R_s}{R_c}\rp\right.
        \nonumber
        \\
        &&\left.+\frac{R_c}{R_s}\lb\frac{1}{3} \lp\frac{R_c}{R_s}\rp^2
            -\frac{3}{2}\lp\frac{R_c}{R_s}\rp + 3\rb - \frac{11}{6}
    \right\},
    \nonumber\\
    \label{eq:re}
\ee
where we define
\be
	B(\chi) \equiv \frac{a}{3} \lp\frac{\mu m_p}{k_{\rm B}}\rp^4 \lp \frac{GM_c}{4} \rp^3 \frac{(1-\chi)^4}{\chi},
\ee
which has units of mass. In practice, we solve Equation~(\ref{eq:re}) numerically to find the surface radius $R_s$ for a given $\chi$, $\Delta M$, $M_c$, and $R_c$. In the limit $R_c\ll R_s$, this simplifies to
\be
	R_s(\chi) \approx R_c \exp \lb \frac{\Delta M}{4\pi B(\chi)}\rb,
	\label{eq:rph}
\ee
so that the surface radius depends exponentially on the amount of mass in the layer. In the opposite limit, when $\Delta M/B(\chi)\ll1$ corresponding to a small envelope mass or $\chi\ll1$, Equation~(\ref{eq:re}) predicts $R_s\approx R_c$. This matches our expectation for when the layer has fully cooled and lost all of its pressure support. 

Given the relations above, for any given Eddington ratio $\chi$, we can solve for all of the properties of the envelope. This corresponds to a snapshot in time during the envelope's evolution, where $\chi$ goes from high to low values from early to late times. The next important question to answer is, what time does this correspond to? We can estimate this by considering the energy of the layer.

The internal energy of the envelope, including both ideal gas and radiation components, is
\be
	E_{\rm int} = \int_{R_c}^{R_s}
    \lp \frac{3}{2}\frac{\rho k_{\rm B}T}{\mu m_p} + aT^4 \rp4\pi r^2 dr.
    \label{eq:eint integral}
\ee
Substituting $T(r)$ and $\rho(r)$ from above, this is integrated to give, to first order in $R_{c}/R_{s}$, 
\be
    E_{\rm int} &\approx& 4\pi a\lp \frac{\mu m_p}{k_{\rm B}} \rp^4
	\lp \frac{GM_c}{4}\rp^4 \frac{(1+\chi)(1-\chi)^4}{2\chi} 
    \nonumber
    \\
    &&\times \frac{1}{R_c} \left[ 1 + \frac{10}{3} \frac{R_c}{R_s} + 4 \frac{R_c}{R_s} \ln \lp \frac{R_c}{R_s} \rp \right] ,
    \label{eq:eint}
\ee
for the internal energy.

Next, there is the gravitational potential energy of the surface layer, which is
\be
	E_{\rm grav} = -\int_{R_c}^{R_s}
     \frac{GM_c}{r} 4\pi r^2\rho dr.
    \label{eq:egrav integral}
\ee
Again substituting $\rho(r)$, we integrate this to find to first order in $R_c/R_s$ that
\be
    E_{\rm grav} &\approx&
    -\frac{16\pi}{3} a\lp \frac{\mu m_p}{k_{\rm B}} \rp^4
	\lp \frac{GM_c}{4}\rp^4 \frac{(1-\chi)^4}{\chi}
    \nonumber
    \\
	&&\times \frac{1}{R_c} 
	\left[ 1 + \frac{3}{2} \frac{R_{c}}{R_{s}} + 3 \frac{R_{c}}{R_{s}} \ln \lp \frac{R_{c}}{R_{s}} \rp \right].
    \label{eq:egrav}
\ee
Comparing this with Equation~(\ref{eq:eint}), in the limit of $R_c\ll R_s$ we see that
\be
    E_{\rm grav} \approx -\frac{8}{3}(1+\chi)^{-1}E_{\rm int}.
    \label{eq:virial}
\ee
In Appendix~\ref{sec:appendix}, we provide a more detailed derivation of the virial relation of a hot layer and a more exact version of Equation~(\ref{eq:virial}). The total energy of the layer is therefore
\be
    E_{\rm tot} = E_{\rm int} + E_{\rm grav}
    \approx -\lb \frac{8}{3}(1+\chi)^{-1}-1\rb E_{\rm int}.
\ee
Energy conservation tells us
$dE_{\rm tot}/dt = -L$, thus
\be
    \lb \frac{8}{3}(1+\chi)^{-1}-1\rb
    \frac{dE_{\rm int}}{dt} = L.
\ee
This expression shows how the internal energy increases with time as the layer compresses and gets hotter.

For the final step, we assume that the internal energy increases like a power law with time as the layer compresses, so that $E_{\rm int}\propto t^{\beta}$. If this is the case, then $dE_{\rm int}/dt = \beta E_{\rm int}/t$. Substituting this into the expression above, we can solve for the time
\be
    t \approx
    \lb \frac{8}{3}(1+\chi)^{-1}-1\rb
    \beta\frac{E_{\rm in}}{L}.
    \label{eq:time}
\ee
As we show below, $\beta\approx1$ is a fairly good approximation, which is what we use for all of our calculation.

The model described above allows us to easily explore the full space of envelope solutions. The main parameters for any given envelope evolution are $M_c$,  $R_c$, and $\Delta M$, and we can explore the evolution by varying $\chi$ and then solving for $t$ using Equation~(\ref{eq:time}). One additional issue is that in a realistic scenario, $R_s$ will start with some value $R_{s,0}$ at early times set by the initial conditions \citep[e.g.,][]{Shen12,Schwab2021}, while Equation~(\ref{eq:rph}) implies the envelope radius can become arbitrarily large. This is because at early times (basically before a Kelvin-Helmholtz cooling timescale), the evolution is still dominated by initial conditions and not in a self-similar regime yet. Thus for times when $R_s(\chi)>R_{s,0}$, we instead use Equation~(\ref{eq:rph}) to solve for $\chi_0$ such that $R_s(\chi_0)=R_{s,0}$, and use this for the envelope profile at these early times.

\subsection{Envelope Profile Evolution}
\label{sec:profiles}

\begin{figure}
\includegraphics[width=0.45\textwidth,trim=0.5cm 5.5cm 2.0cm 3.0cm]{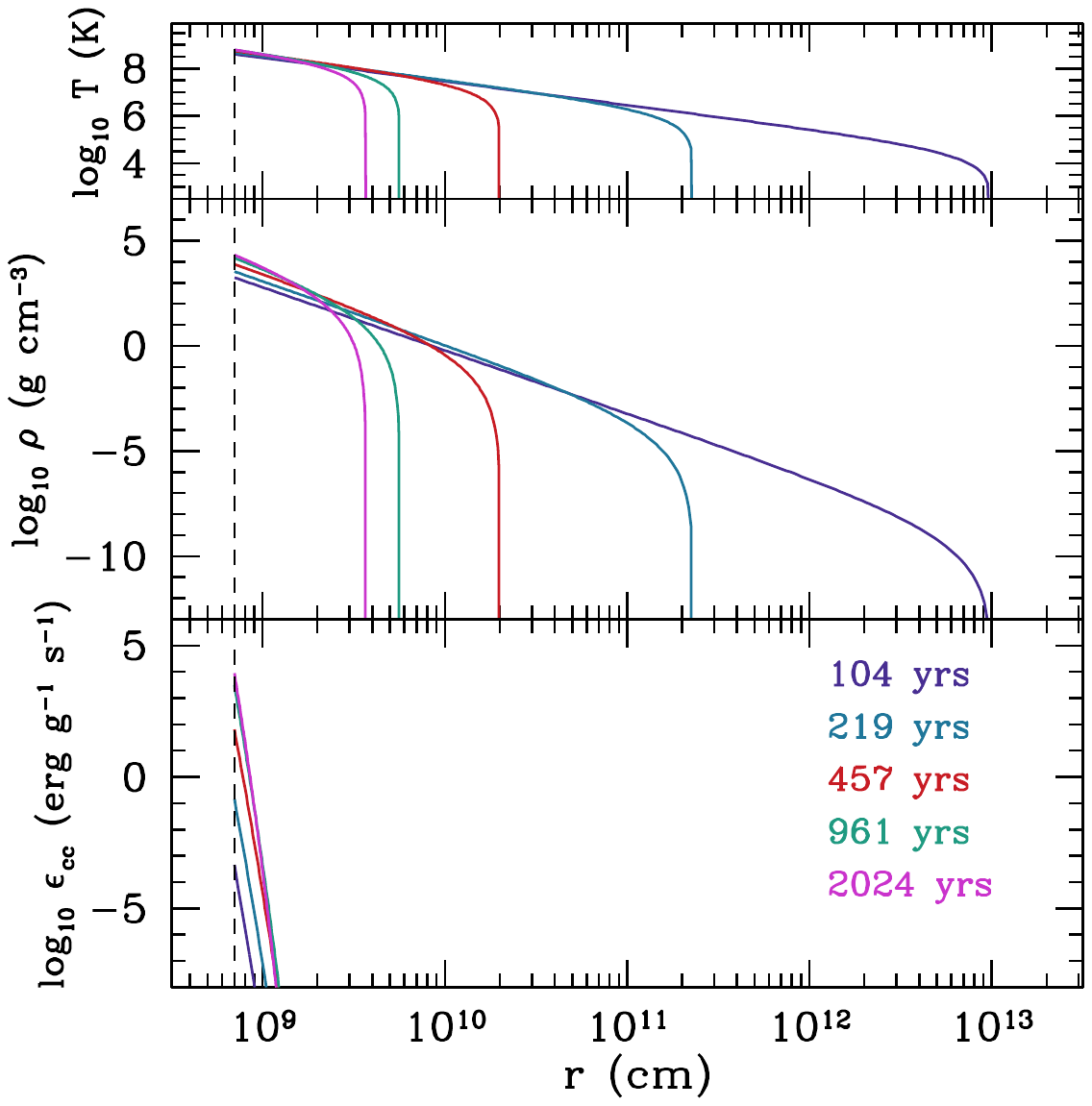}
\caption{Profiles of $T$, $\rho$ and the carbon-carbon burning rate $\epsilon_{\rm CC}$ for a cooling envelope model with $M_c=1.25\,M_\odot$, $\Delta M=0.03\,M_\odot$, and $R_c=7\times10^8\,{\rm cm}$. Each color line represents a different $\chi$ (see text for the specific values), which in turn corresponds to a different time during the cooling as shown in the bottom panel. The vertical black dashed line represents the surface of the core.}
\label{fig:profiles}
\end{figure}

We begin by producing an example series of envelope profiles for different values of $\chi$, which are summarized in Figure~\ref{fig:profiles}. This displays the profiles for the temperature, density, and carbon burning rate $\epsilon_{\rm CC}$ (which we discuss in more detail in the Section~\ref{sec:c12}). For this example, we set $M_c=1.25\,M_\odot$, $\Delta M=0.03\,M_\odot$, and $R_c=7\times10^8\,{\rm cm}$. We set the opacity to $\kappa=0.2\,{\rm cm^2\,g^{-1}}$ representing highly ionized intermediate mass elements, and $\mu=1.7$ as appropriate for equal mass fractions of C and O. To vary the time, we use values of $\chi=0.665$, $0.612$, $0.532$, $0.432$, and $0.370$ from early to late times. The initial $\chi$ is set to give an initial radius of $R_s\approx10^{13}\,{\rm cm}$, similar to merger remnant studies \citep[e.g.,][]{Schwab12}. The corresponding times using Equation~(\ref{eq:time}) are shown in the bottom panel of Figure~\ref{fig:profiles}.

This series of profiles demonstrates how the hot layer contracts from $\approx10^{13}\,{\rm cm}$ to $\approx4\times10^9\,{\rm cm}$ over a timescale of $2,024\,{\rm yrs}$. Prior to the first profile shown at $104\,{\rm yrs}$, the layer does not contract appreciably and remains at roughly $10^{13}\,{\rm cm}$. This $104\,{\rm yr}$ is roughly the Kelvin-Helmholtz timescale of the initial configuration. Changing this choice of initial radius will change the initial evolution, but does not alter the ensuing self-similar evolution for most times.

As the layer contracts, the base temperature and density of the layer increase. Although it may be difficult to resolve since the plot shows such a large dynamic range, in the earliest model has a base temperature and density of $4.1\times10^8\,{\rm K}$ and $1.8\times10^3\,{\rm g\,cm^{-3}}$, respectively, while the final model has a base temperature and density of $6.2\times10^8\,{\rm K}$ and $2.1\times10^4\,{\rm g\,cm^{-3}}$, respectively. The carbon-burning rate increases by $7$ orders of magnitude at the base!

\subsection{Evolution Dependence on $R_c$, $M_c$, and $\Delta M$}
\label{sec:dependence}

The main observables for the central star of Pa~30 are the time since the SN, the current luminosity, and the photospheric radius. Thus it is helpful to vary each of the parameters in this model independently to explore how these observables are expected to change.

\begin{figure}
\includegraphics[width=0.45\textwidth,trim=0.5cm 5.5cm 2.0cm 3.0cm]{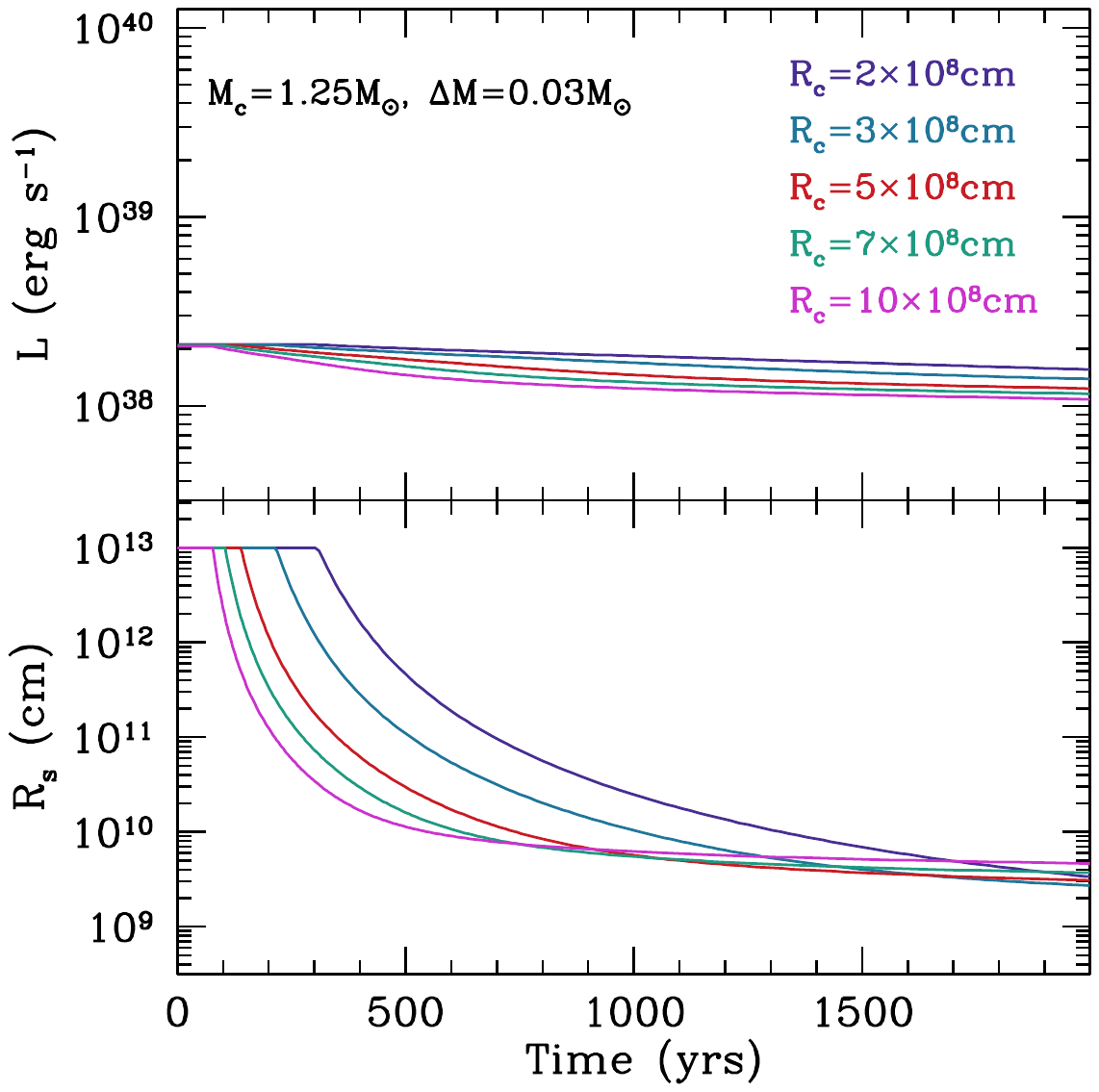}
\caption{Time evolution of the luminosity $L$ and the envelope radius $R_s$, each keeping $M_c=1.25\,M_\odot$ and $\Delta M=0.03\,M_\odot$ fixed, while varying $R_c$ with values indicated by the different colored lines. This demonstrates that, over the timescales of interest, $L$ varies only slightly, with a larger $L$ for smaller $R_c$. The radius $R_s$ depends very sensitively to $R_c$, with an inverse relationship between the two (although at sufficiently late times these lines cross and a smaller $R_c$ will asymptote to a smaller $R_s$).}
\label{fig:rcore}
\end{figure}

We start in Figure~\ref{fig:rcore} by fixing $M_c=1.25\,M_\odot$ and $\Delta M=0.03\,M_\odot$ and varying $R_c$ over a range of radii that may be expected for a WD. Changing the core radius has a fairly little impact on the luminosity, although the changes do grow at later times. The impact of $R_c$ on $R_s$ is more noticeable. First, the initial evolution, with $R_s$ roughly fixed, lasts longer for smaller $R_c$. Second, during the first few thousand years $R_s$ is larger when $R_c$ is smaller (although at sufficiently late times the models cross and $R_s\approx R_s$). The reason for this inverse relationship between $R_s$ and $R_c$ is that a smaller $R_c$ implies a larger internal energy because the gravitational binding energy of the surface layer is  larger. Thus, a layer with a small $R_c$ takes more time to cool, and the radius $R_s$ remains larger longer.

\begin{figure}
\includegraphics[width=0.45\textwidth,trim=0.5cm 5.5cm 2.0cm 3.0cm]{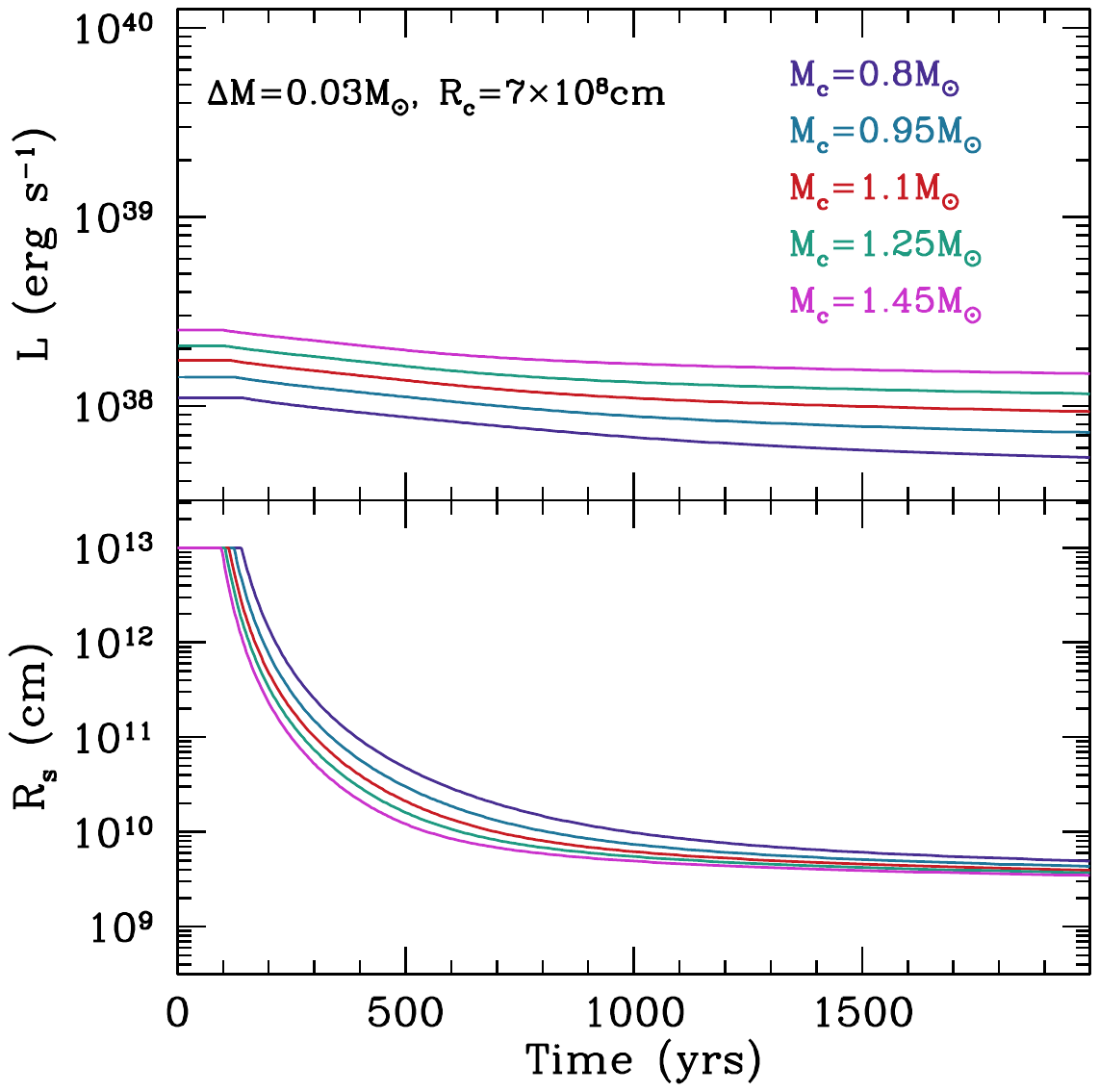}
\caption{Similar to Figure~\ref{fig:rcore}, but this time fixing $\Delta M=0.03\,M_\odot$ and $R_c=7\times10^8\,{\rm cm}$, while varying $M_c$. Here, the main change is in the luminosity, with larger $M_c$ corresponding to a larger $L$. Since the dynamic range of possible $M_c$ values is relatively small, the impact on the $R_s$ evolution is not very strong.}
\label{fig:mcore}
\end{figure}

\begin{figure}
\includegraphics[width=0.45\textwidth,trim=0.5cm 5.5cm 2.0cm 3.0cm]{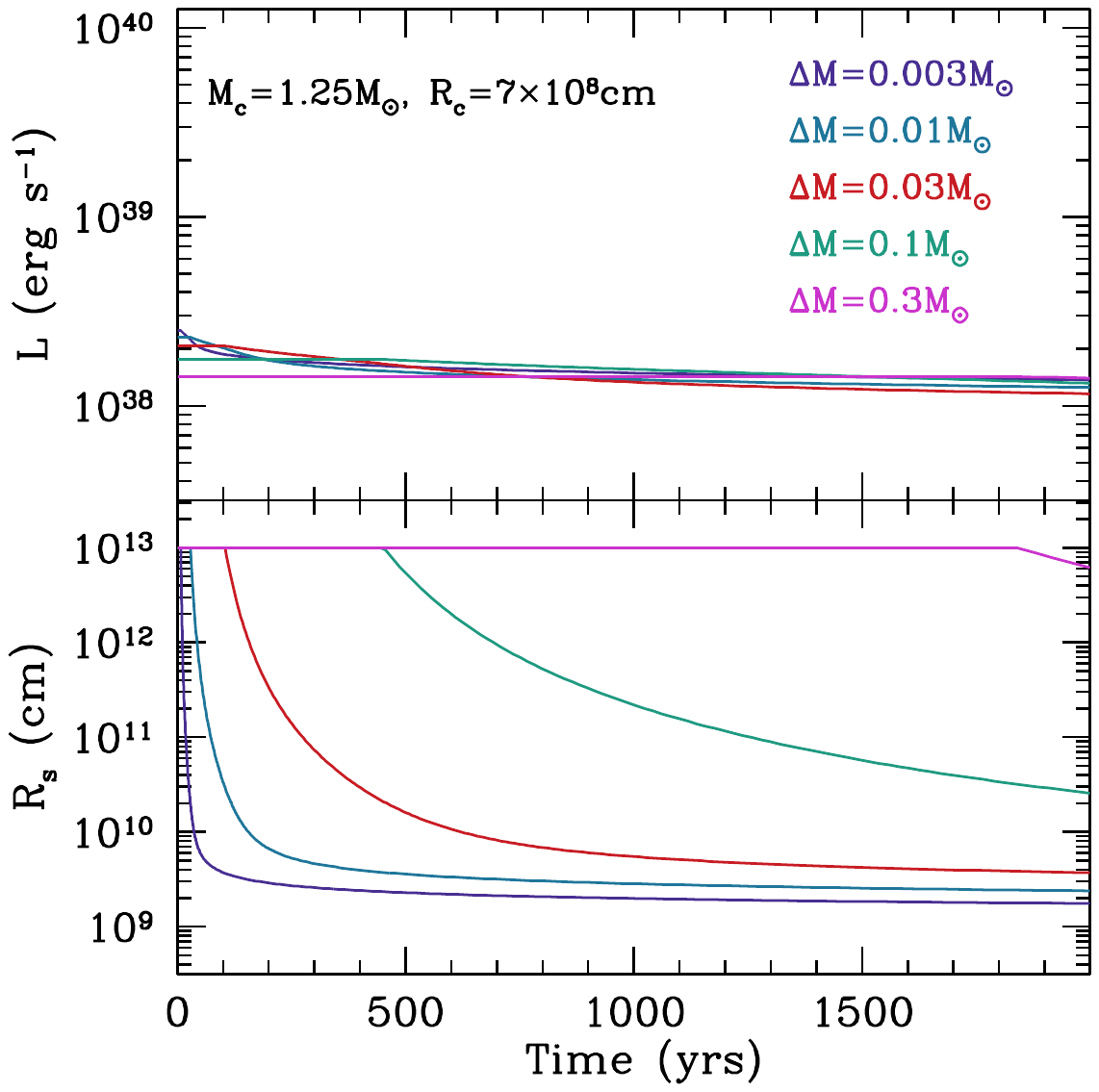}
\caption{Similar to Figure~\ref{fig:rcore}, but this time fixing $M_c=1.25\,M_\odot$ and $R_c=7\times10^8\,{\rm cm}$, while varying $\Delta M$. Here we see that the observed radius can vary dramatically with envelope mass, and at sufficiently large $\Delta M$ it does not change appreciably over the observational time period.}
\label{fig:deltam}
\end{figure}

In Figure~\ref{fig:mcore}, we fix $\Delta M$ and $R_c$ and vary $M_c$. The luminosity is larger for a larger $M_c$ (and vice versa), while the radius evolution is roughly unchanged. We conclude that $M_c$ is best constrained by the current observed luminosity.

Finally, in Figure~\ref{fig:deltam}, we fix $M_c$ and $R_c$ and vary $\Delta M$. We see that $R_s$ can change very strongly depending on the value of $\Delta M$, which makes sense because of the exponential dependence of the radius on the mass as shown by Equation~(\ref{eq:rph}). Although note that we consider a much larger dynamic range for $\Delta M$ in comparison to the other parameters because there is more uncertainty in what its value could be.

\begin{figure}
\includegraphics[width=0.45\textwidth,trim=1.0cm 5.5cm 2.0cm 3.0cm]{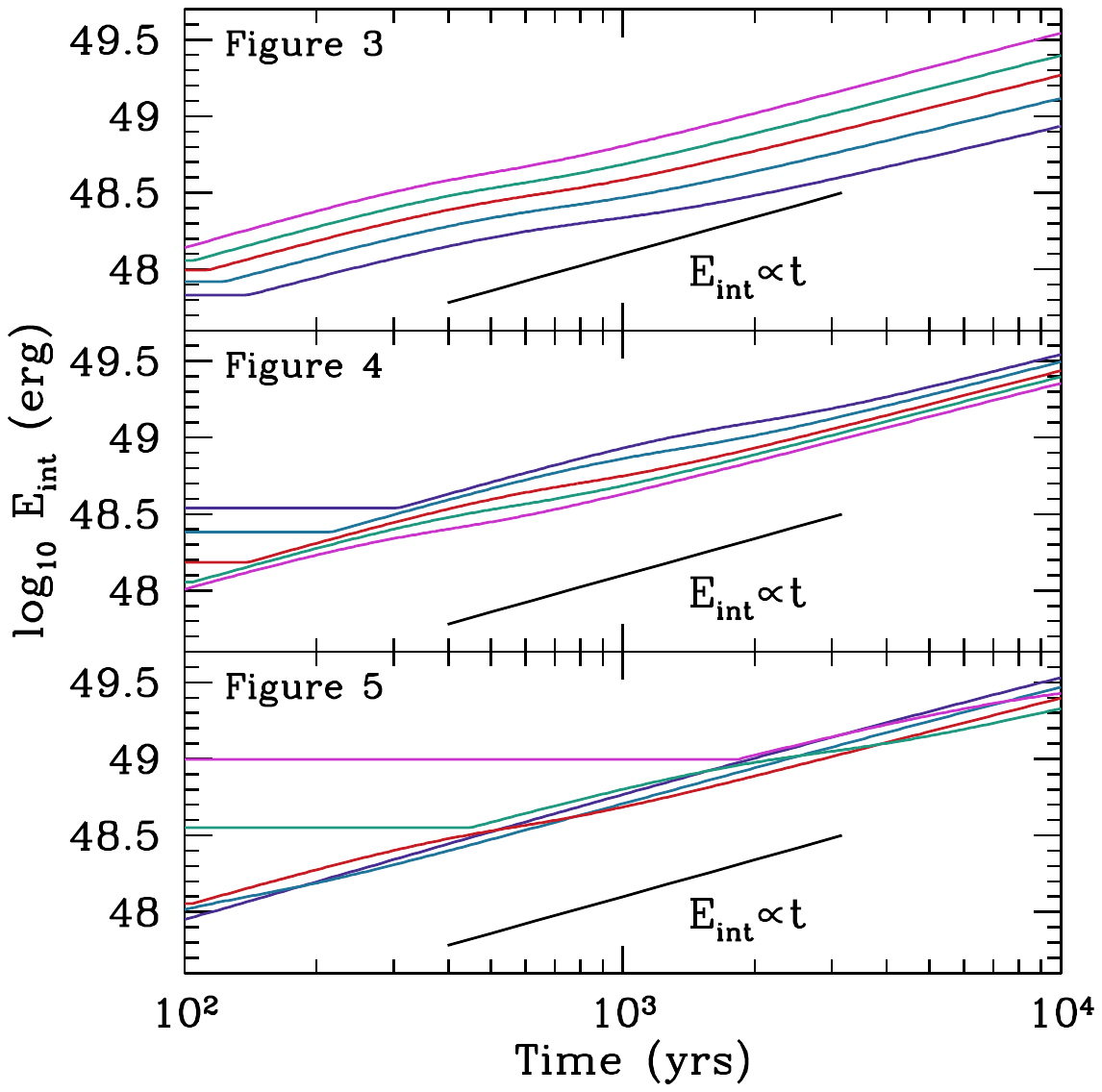}
\caption{The evolution of the internal energy $E_{\rm int}$ versus time using Equations~(\ref{eq:eint}) and  (\ref{eq:time}). The 15 models shown match the models from Figures~\ref{fig:rcore}, \ref{fig:mcore}, and \ref{fig:deltam} as labeled with the line colors matching the corresponding models. The important point is that plotting this evolution logarithmically shows that $\beta\approx1$ is a good approximation as used throughout this work for the calculations.}
\label{fig:energy check}
\end{figure}

An important factor when setting the time $t$ for the evolution is the internal energy's power law dependence on the time $\beta$, as shown by Equation~(\ref{eq:time}). To check this, in Figure~\ref{fig:energy check} we plot the time evolution of $E_{\rm int}$ for all 15 models from Figures~\ref{fig:rcore}, \ref{fig:mcore}, and \ref{fig:deltam}. We intentionally plot both axis logarithmically to focus on the power law dependence. In comparison, we also include a line representing $E_{\rm int}\propto t$ (for $\beta=1$). This demonstrates that setting $\beta\approx 1$ for this work is a fairly good approximation, although it can introduce roughly $\sim10\%$ uncertainties in the time estimate.

\section{Carbon Burning}
\label{sec:c12}

As the C/O-rich envelope contracts, carbon burning can turn on abruptly, and act as an additional heat source that slows contraction and helps maintain a near-Eddington luminosity \citep[e.g.,][]{Kashiyama19,Ko2026}. Here, we explore which models are expected to initiate appreciable carbon burning at the base of the envelope and over what timescales.

For the C-C burning energy generation rate, we use \citep{Caughlan1988,Kippenhahn2013}
\be
    \epsilon_{\rm CC}
    &&= 1.86\times10^{43}
    X_{12}^2T_9^{-3/2}T_{9a}^{5/6}
    \nonumber
    \\
    &&
    \times\exp[-84.165/T_{9a}^{1/3}-2.12\times10^{-3}T_9^3]
    \,{\rm erg\,s^{-1}\,g^{-1}},
    \nonumber
    \\
    \label{eq:carbon burning}
\ee
where $X_{12}$ is the carbon mass fraction, $T_9=T/10^9\,{\rm K}$, and $T_{9a}=T_9/(1+0.0396T_9)$. We do not include screening corrections due to the relatively low densities in the hot envelope, and we simply set $X_{12}=0.5$. Equation~(\ref{eq:carbon burning}) is used to plot the profile of carbon burning energy generation rate in Figure~\ref{fig:profiles}, demonstrating how strongly peaked the burning rate is at the base of the envelope due to its temperature sensitivity. 

For each of our models, we also numerically integrate the carbon burning rate to get the total luminosity from carbon burning,
\be
    L_{\rm CC}
    = \int_{R_c}^{R_s}
    4\pi r^2 \rho \epsilon_{\rm CC} dr.
    \label{eq:carbon luminosity}
\ee
We expect that once $L_{\rm CC}\approx L$, then carbon burning should pause the contraction of the hot envelope due on a timescale of how long it takes to burn a significant fraction of the carbon in the layer (which would be much longer than the age of Pa~30).

\begin{figure}
\includegraphics[width=0.45\textwidth,trim=1.0cm 5.5cm 2.0cm 3.0cm]{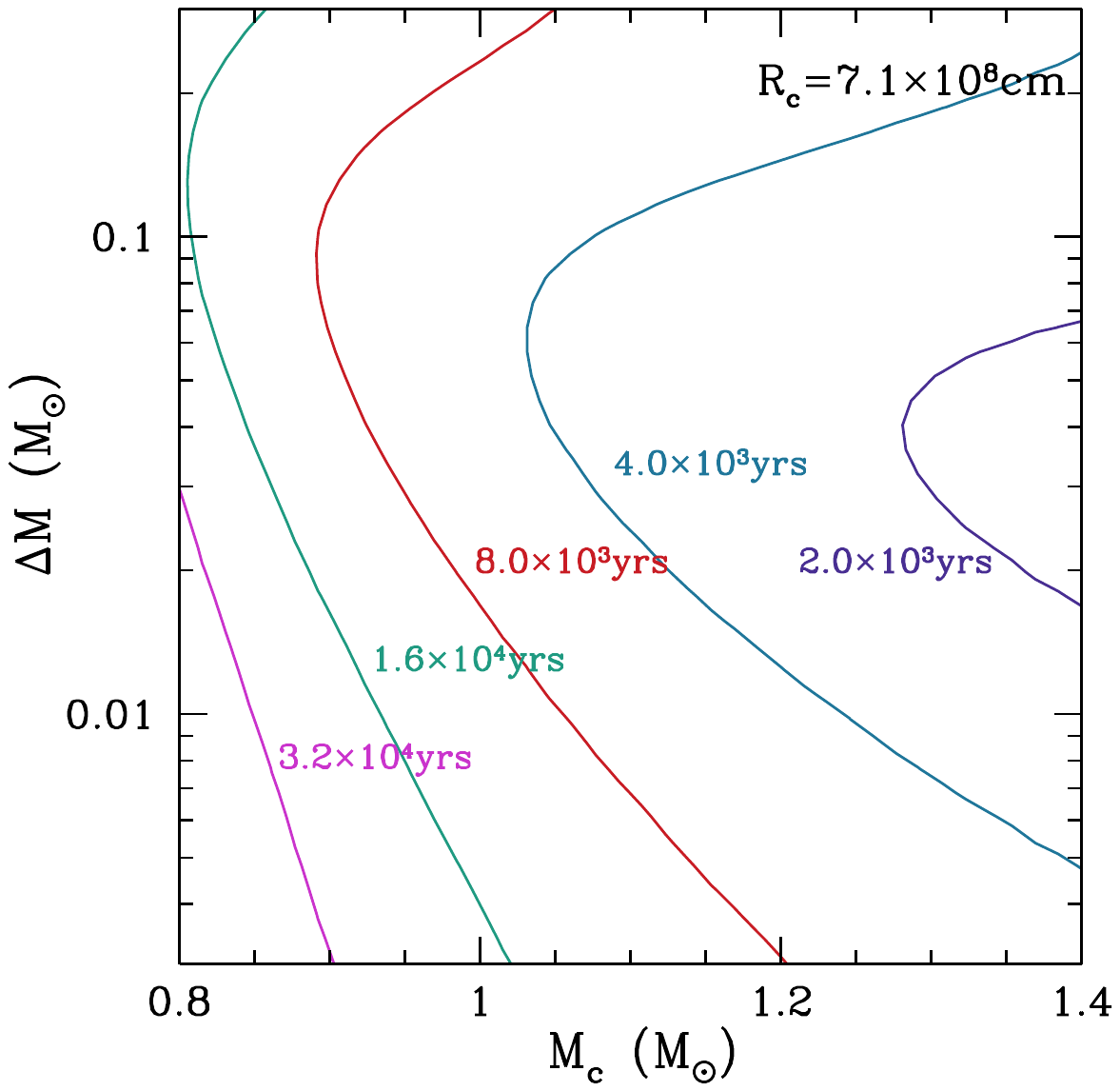}
\caption{Contours show the time it takes to reach $L_{\rm CC}\approx L$ as a function of $M_c$ and $\Delta M$ for fixed $R_c$. This shows the general trend that a larger $M_c$ increases the surface gravity and base temperature, leading to earlier carbon burning. The dependence on $\Delta M$ is more complex, exhibiting a ``sweet spot'' at $\Delta M\approx 0.03-0.1\,M_\odot$, beyond which the burning time increases. We discuss this behavior in further detail in the text.}
\label{fig:fixed rcore}
\end{figure}

\begin{figure}
\includegraphics[width=0.45\textwidth,trim=1.0cm 5.5cm 2.0cm 3.0cm]{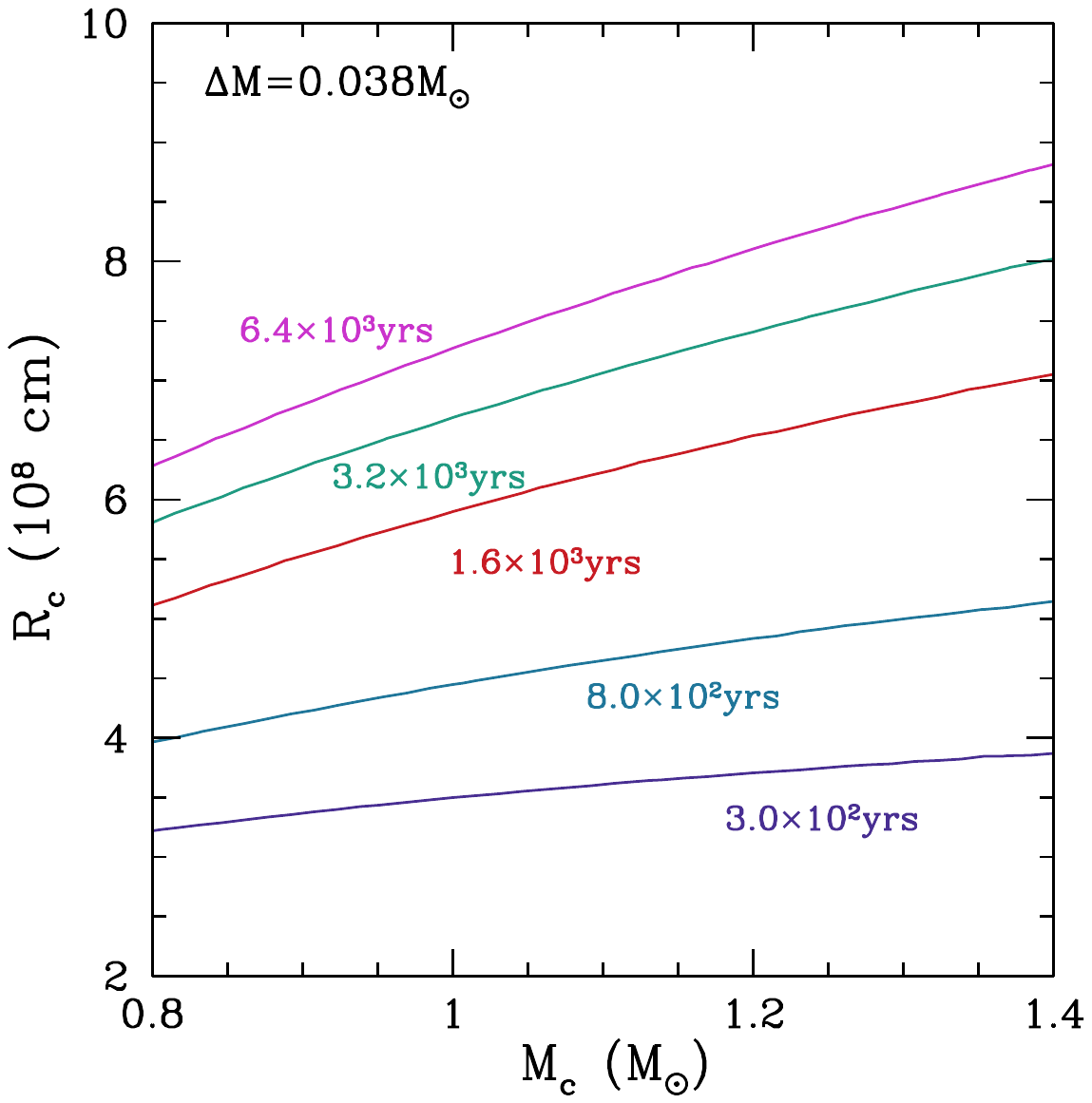}
\caption{Similar to Figure~\ref{fig:fixed rcore}, but now showing when $L_{\rm CC}\approx L$ as a function of $M_c$ and $R_c$ for fixed $\Delta M$. This shows that the burning timescale depends very sensitively on $R_c$. A more compact core has a much hotter base temperature.}
\label{fig:fixed deltam}
\end{figure}

To better quantify when we expect carbon burning to become important, in Figures~\ref{fig:fixed rcore} and \ref{fig:fixed deltam} we plot contours of the time it takes to reach $L_{\rm CC}\approx L$. In Figure \ref{fig:fixed rcore}, increasing $M_c$ leads to earlier carbon ignition because of the larger gravity and correspondingly large base temperature for the hot layer. The dependence on $\Delta M$ is more subtle. At high $\Delta M$, the contraction of the envelope is slower because the cooling timescale of the layer becomes longer. At low $\Delta M$, the density at the base of the layer becomes smaller, again leading to a longer time to ignite carbon. The result of these effects is that there is an in-between region, roughly $\Delta M\approx 0.03-0.1\,M_\odot$ depending on $M_c$, where carbon ignites fastest.

In Figure~\ref{fig:fixed deltam}, we fix $\Delta M$ and vary the other parameters. The time contours in this case are simpler, indicating that the time to carbon ignition is primarily sensitive to $R_c$. This is because when the core is more compact, the layer is deeper in the gravitational potential well and has a hotter base temperature. Thus, determining whether burning occurs can be a sensitive constraint on the core's thermal state.

\begin{figure}
\includegraphics[width=0.45\textwidth,trim=0.5cm 5.5cm 2.0cm 3.0cm]{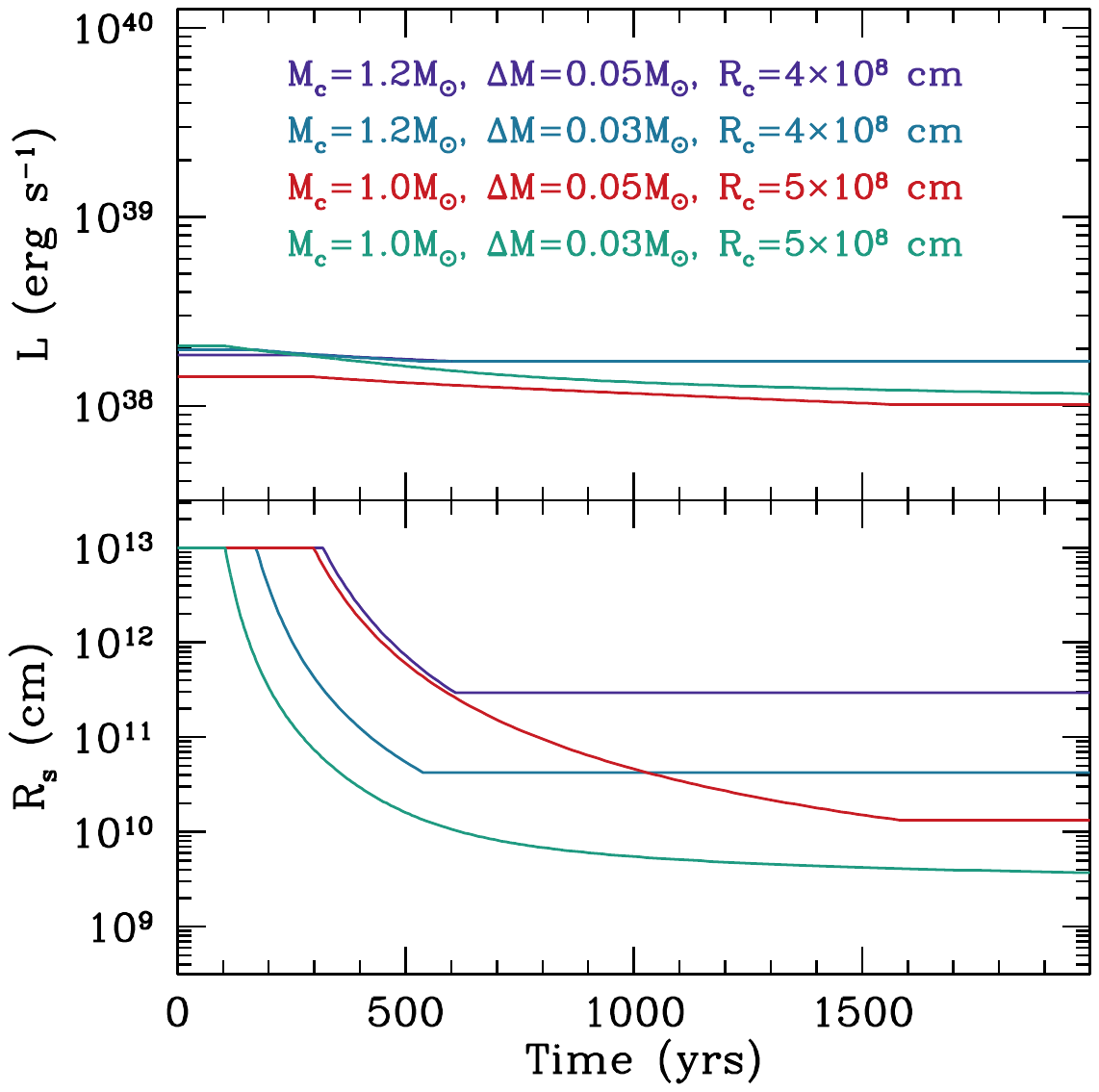}
\caption{Time evolution of the luminosity $L$ and the envelope radius $R_s$ for example models that ignite carbon. When $L_{\rm CC}=L$, we assume that the luminosity and radius $R_s$ stay fixed throughout the remainder of the evolution. The specific parameters used are summarized in the upper panel.}
\label{fig:burning}
\end{figure}

In cases where carbon does ignite, we assume that the luminosity and radius $R_s$ remain fixed for the remainder of the evolution that we are following. This will continue only as long as sufficient carbon fuel remains, but for the timescales we are more interested in, we do not follow the subsequent evolution after the carbon is depleted. Examples of this evolution are shown in Figure~\ref{fig:burning}, where the specific parameters were simply chosen to show some diversity in evolution.

An important takeaway from this exploration is that carbon burning is not a necessity to explain the thermal state of Pa~30 today. If the layer has a sufficiently small or large envelope mass, or if the core radius is sufficiently large, then we do not expect carbon burning to ignite during the $\approx845\,{\rm yrs}$ that Pa~30 has been evolving. Conversely, as more detailed observations are made of Pa~30, and if they indeed show (through the nucleosynthetic inferences, for example) that burning is necessary, then these types of models will help narrow the parameter space where this is possible.


\section{Constraints on Pa~30}
\label{sec:pa30}

Given the discussion above of how the evolution depends on $M_c$, $R_c$, and $\Delta M$, we next seek to constrain the values inferred from the current properties of Pa~30 of an age of $\approx845\,{\rm yrs}$, a luminosity of $\approx1.5\times10^{38}\,{\rm erg\,s^{-1}}$, and a photospheric radius $\approx 0.15\,R_\odot$ \citep{GvaramadzeV+19_Nat}. Although we use these values for the fitting, this comes with some caveats. As noted in \citet{Schaefer23}, the full spectral energy distribution of the central star in Pa~30 is not measured, so inferences must be made about the bolometric luminosity. Given that we see a dense wind \citep[e.g.,][]{Kashiyama19}, the observed photosphere may represent a radius within the wind and thus the true stellar photosphere is smaller. The advantage of the framework we present here is that, as these specific values are updated with future measurements and models, we can readily assess how the underlying physical properties of the central star change.

To explore what models fit Pa~30 best, we run a grid of $10^6$ models varying $M_c$, $R_c$, and $\Delta M$. We resolve the full time evolution for each one so that we can track the luminosity of carbon burning at each time using Equation~(\ref{eq:carbon luminosity}). If at some time $L_{\rm CC}=L$, we assume that the luminosity stops evolving and continue at a constant level for the remainder of time (as shown in the examples in Figure~\ref{fig:burning}). For each of these models, we then extract the parameters at $t=845\,{\rm yrs}$ and calculate a fitting parameter
\be
    \chi^2= \lp 1-R_s/R_{s,\rm obs}\rp^2
    + \lp 1-L/L_{\rm obs} \rp^2,
    \label{eq:chi2}
\ee
where $R_{s,\rm obs}$ and $L_{\rm obs}$ are the observed photospheric radius and luminosity, respectively. This $\chi^2$ is meant to represent the goodness of fit while keeping both the radius and luminosity on roughly equal footing.

Calculating $\chi^2$ across all $10^6$ models that we run, we identify the model with the smallest $\chi^2$ value, which is $M_c=1.3\,M_\odot$, $\Delta M=0.038\,M_\odot$, and $R_c=7.1\times10^8\,{\rm cm}$. These values have interesting implications. First, $M_c$ is on the higher side for what may be expected for a WD. This is in line with arguments that the primary WD in the event that made SN~1181 was probably an O/Ne WD rather than a C/O WD \citep[where a typical boundary between these compositions is $\sim1.05\,M_\odot$,][]{Siess2006,Doherty2015}. Next, the amount of fallback and other burnt material still left in the hot surface layer is relatively small. Combining this with the \mbox{$\approx0.5\,M_\odot$} of ejecta inferred from modeling the SN remnant X-rays \citep{KoT+24}, argues that the secondary mass was $\approx0.5-0.6\,M_\odot$, consistent with a typical C/O WD. Finally, the relatively large $R_c$ we measure indicates that significant heat entered the core during the merger and the subsequent deflagration that produced SN~1181. These values we find for $M_c$, $R_c$, and $\Delta M$ are roughly consistent with what is found by \citet{Ko2026}. For their approach, they consider the temperature of the core rather than its radius, but we both agree that the core must be hotter than a zero-temperature WD.

\begin{figure}
\includegraphics[width=0.45\textwidth,trim=0.5cm 5.5cm 2.0cm 3.0cm]{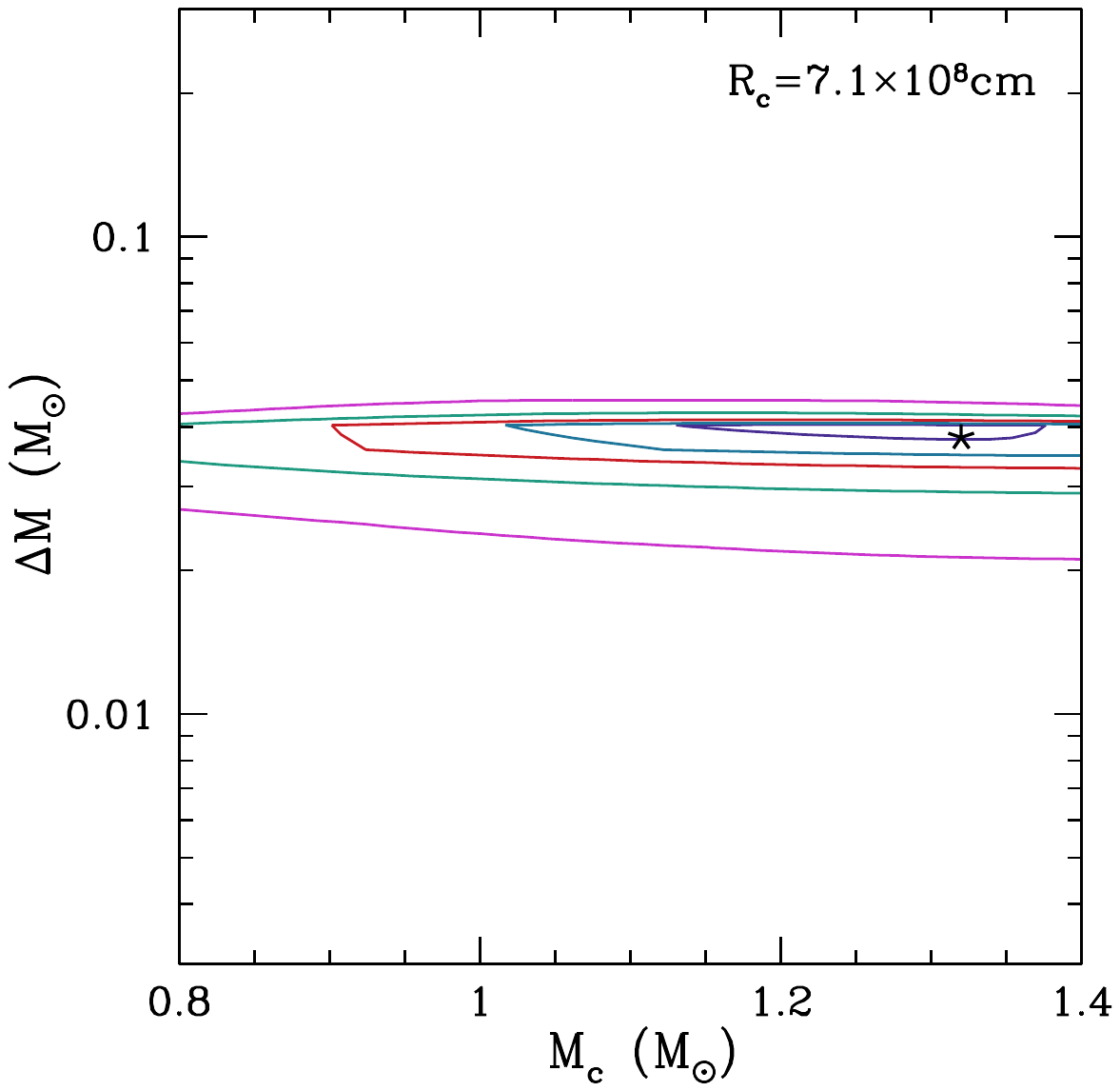}
\caption{Contours of constant $\chi^2$ found when setting $R_c=7.1\times10^8\,{\rm cm}$ and varying $M_c$ and $\Delta M$. From inside out, the contours represent values of $\chi^2=0.02$, $0.04$, $0.08$, $0.16$, and $0.32$. The star represents the best-fit values.}
\label{fig:fixed rcore2}
\end{figure}

\begin{figure}
\includegraphics[width=0.45\textwidth,trim=1.0cm 5.5cm 2.0cm 3.0cm]{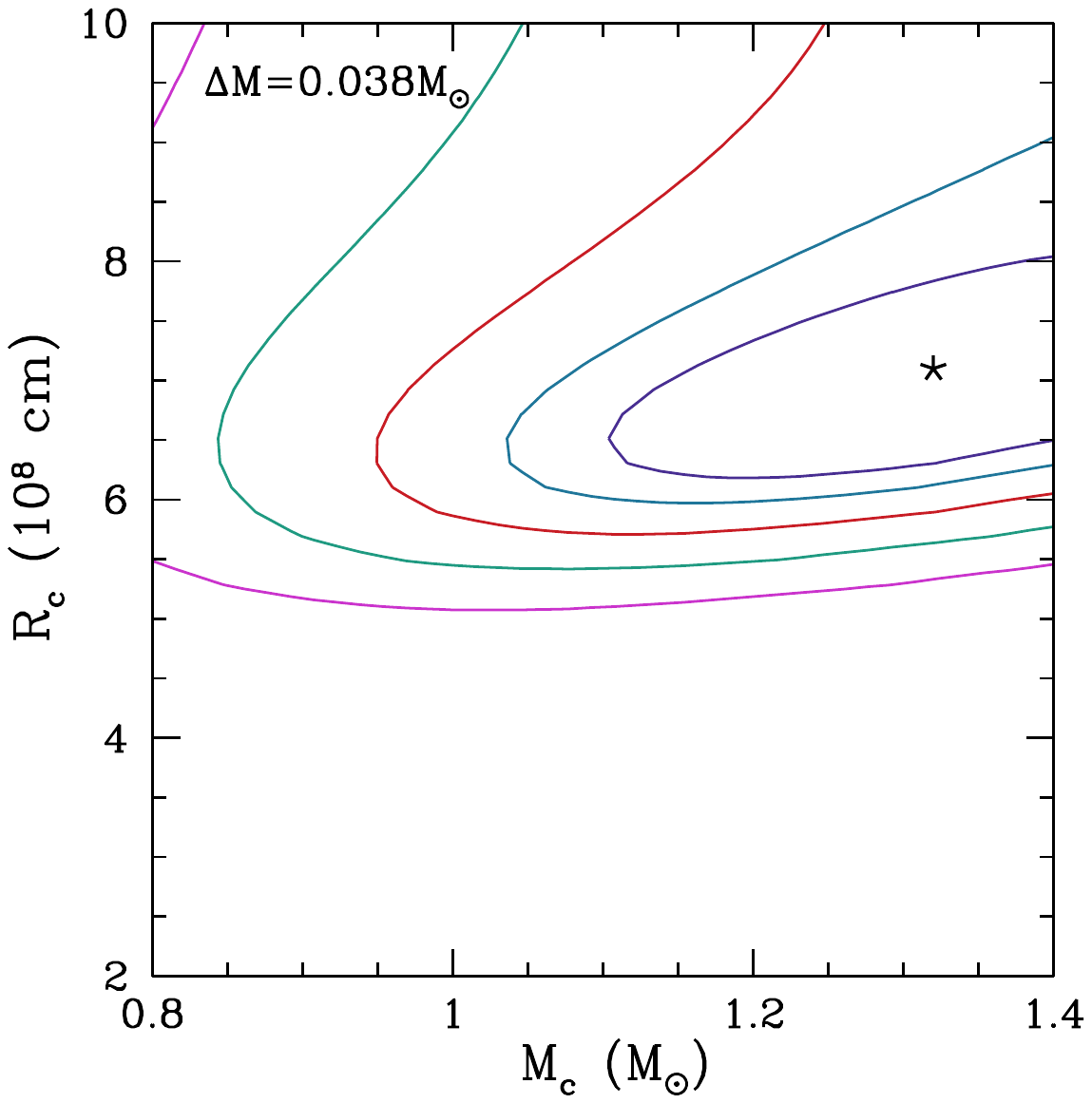}
\caption{Similar to Figure~\ref{fig:fixed rcore2}, but now with fixed $\Delta M=0.038\,M_\odot$ and varying $M_c$ and $R_c$. From inside out, the contours represent values of $\chi^2=0.01$, $0.02$, $0.04$, $0.08$, and $0.16$. The star represents the best fit values.}
\label{fig:fixed deltam2}
\end{figure}

To better understand how well these parameters are constrained, in Figure~\ref{fig:fixed rcore2} we fix $R_c$ at its best-fit value and show contours of constant $\chi^2$. The star indicates the best fit value of $M_c$ and $\Delta M$. We see that values of $M_c\approx1.15-1.4\,M_\odot$ are roughly equal fits. Thus, $M_c$ is large, but there remains some uncertainty in its exact value. On the other hand, $\Delta M\approx0.02-0.04$ and is fairly well constrained to be small. This is related to our discussion in Section~\ref{sec:dependence}, which demonstrated that the extent of the envelope was very sensitive to $\Delta M$. In other words, $\Delta M$ must be fairly small for Pa~30 to be so compact in a relatively short amount of time.

In Figure~\ref{fig:fixed deltam2}, we fix $\Delta M$ and explore the different values of $\chi^2$ as a function of $M_c$ and $R_c$. Here we see that there is some degeneracy between these two parameters, but generally large values are favored, $M_c\approx1.15-1.4\,M_\odot$ and $R_c\approx(6-8)\times10^8\,{\rm cm}$, consistent with a hot O/Ne core.

\section{Discussion and Future Work}
\label{sec:conclusion}

We explored the thermal evolution of the central star of SN~1181, which now sits at the center of Pa~30. We approximate it as a hot envelope sitting above a colder core. Using a semi-analytic model, we show that the thermal evolution of such a mass configuration mainly depends on three parameters, the mass of the core $M_c$, the radius of the core $R_c$, and the mass of the hot envelope $\Delta M$. Our semi-analytic framework simplifies our ability to explore the range of evolutions for different sets of parameters. We also discussed carbon burning within the layer and when it is expected to occur.

We fit our model to the current age and state of the Pa~30 central star, and we conclude that the envelope needs to be relatively low mass ($\Delta M\approx0.02-0.04\,M_\odot$) for it to contract to the size that is observed today.
Combining our small envelope mass with the observed mass for the supernova remnant of $\approx0.5\,M_\odot$ \citep{KoT+24} suggests a C/O WD secondary that was mostly ejected in the explosion. There is some degeneracy in the core mass and radius, but the favored values of $M_c\approx1.15-1.4\,M_\odot$ and $R_c\approx(6-8)\times10^8\,{\rm cm}$ are consistent with an O/Ne WD primary that is somewhat expanded by heating during the merger and explosive event (deflagration) that produced SN~1181. Our best-fit model reproduces the current luminosity of the central star solely through thermal emission, without invoking carbon burning.

We acknowledge that this model has important caveats. For example, there are uncertainties in the true bolometric luminosity \citep[e.g.,][]{Schaefer23} and the true radius given the dense wind \citep[e.g.,][]{Kashiyama19}. The strength of our approach, however, is that we can readily understand how the system's basic parameters change as these observed properties are updated.

An important piece of physics that is missing from this model is how the radius adjusts when carbon ignites, since this presumably would generate convection, alter the envelope's density profile, and require tracking the envelope's entropy, all of which are outside the scope of our framework. We plan to address this shortcoming in future work using full numerical stellar models \citep[although see the investigation by][]{Ko2026}. Our framework also does not include contributions to the pressure in the hot envelope from nonrelativistic degenerate electrons, although we consider this a small correction at these high temperatures.

Going forward, it will be important to understand how the merger, unstable ignition, and subsequent deflagration produced the mass distribution for the central star in Pa~30 observed today. While this will require detailed numerical simulations, the subsequent time evolution of such models can be addressed quickly using the intuition we have developed from the framework presented here. This may also help predict the diversity of hot stars left from SNe~Iax that may exist elsewhere in the Milky Way or nearby galaxies.

\acknowledgments

We thank Ilaria Caiazzo, Tim Cunningham, Jim Fuller, Ken Shen, and Daichi Tsuna for helpful discussions. YZ acknowledges support from MAOF grant 12641898 and visitor support from the Observatories of the Carnegie Institution for Science, Pasadena CA, where much of this work was completed.

\begin{appendix}
\counterwithin{figure}{section}

\section{Virial Relation for a Hot Surface Layer}
\label{sec:appendix}

An important step in deriving the time in Equation~(\ref{eq:time}) is understanding the relation between the internal and gravitational energy, i.e., the virial relation. Typically one might assume that $E_{\rm grav}=-2E_{\rm int}$, but this is inconsistent with what we find in Equation~(\ref{eq:virial}) from direct integration of the energies for a constant luminosity profile. Here we present an alternative derivation that yields the same relation as Equation~(\ref{eq:virial}), thereby showing that this result is indeed robust.

Starting with hydrostatic balance, $dP/dr=-\rho GM_c/r^2$, multiplying both sides by $4\pi r^3$, and integrating over the hot surface layer,
\be
    \int_{R_c}^{R_s} 4\pi r^3 \frac{dP}{dr}dr
    =
    -\int_{R_c}^{R_s} \rho\frac{GM_c}{r}4\pi r^2dr.
\ee
The right side is simply the gravitational potential energy $E_{\rm grav}$. The left side can be integrated by parts,
\be
    -4\pi R_c^3P(R_c) - 3\int_{R_c}^{R_s} 4\pi r^2 P dr
    = E_{\rm grav}.
    \label{eq:relation}
\ee
The term on the far left side is a key difference between a hot layer and a full self-gravitating body like a star. In the latter case
\be
    \left.4\pi r^3 P\right|_{R_c}^{R_s} = 0,
\ee
because $R_c=0$ and $P(R_s)\approx 0$. For the hot layer this term is non-zero, which we show ultimately leads to a modified virial relation.

Pressure can be related to energy by
\be
    \int_{R_c}^{R_s} 4\pi r^2 P dr
    = \frac{2}{3}E_{\rm gas} + \frac{1}{3}E_{\rm rad},
\ee
where $E_{\rm gas}$ and $E_{\rm rad}$ are the contributions to the internal energy from ideal gas and radiation, respectively. Using the analytic envelope solution in the limit  $R_c\ll R_s$, the base-pressure term satisfies so Equation~(\ref{eq:relation}) reduces to
\be
    -\frac{8}{3} E_{\rm gas}
    -\frac{4}{3} E_{\rm rad}
    \approx E_{\rm grav}.
    \label{eq:relation2}
\ee
To find the energies, we integrate them directly from our profile
\be
	E_{\rm gas} = \int_{R_c}^{R_s}\frac{3}{2}
        \frac{\rho k_{\rm B}T}{\mu m_p} 4\pi r^2 dr
    \approx 4\pi a\lp \frac{\mu m_p}{k_{\rm B}} \rp^4
	\lp \frac{GM_c}{4}\rp^4
    \frac{(1-\chi)^5}{2\chi}
    \frac{1}{R_c} 
	\left[ 1 + \frac{10}{3} \frac{R_{c}}{R_{s}} + 4 \frac{R_{c}}{R_{s}} \ln \left( \frac{R_{c}}{R_{s}} \right) \right],
    \label{eq:egas}
\ee
and
\be
	E_{\rm rad} = \int_{R_c}^{R_s}
        aT^4 4\pi r^2 dr
    \approx 4\pi a\lp \frac{\mu m_p}{k_{\rm B}} \rp^4
	\lp \frac{GM_c}{4}\rp^4
    (1-\chi)^4
    \frac{1}{R_c} 
	\left[ 1 + \frac{10}{3} \frac{R_{c}}{R_{s}} + 4 \frac{R_{c}}{R_{s}} \ln \left( \frac{R_{c}}{R_{s}} \right) \right],
    \label{eq:erad}
\ee
where we only keep terms up to first order in the ratio $R_c/R_s$. Comparing these with $E_{\rm int}$ from Equation~(\ref{eq:eint}), we see that
\be
    E_{\rm gas} = \lp \frac{1-\chi}{1+\chi}\rp E_{\rm int}
\ee
and
\be
    E_{\rm rad} = \lp \frac{2\chi}{1+\chi}\rp E_{\rm int}.
\ee
This matches our expectation that when $\chi=0$ then $E_{\rm int} = E_{\rm gas}$ and conversely when $\chi=1$ then $E_{\rm int}=E_{\rm rad}$. Finally, substituting these into Equation~(\ref{eq:relation2}), we find
\be
    E_{\rm grav} 
    \approx -\frac{8}{3}(1+\chi)^{-1} \left[ \frac{ 1 + \frac{10}{3} (R_{c}/R_{s}) + 4 (R_{c}/R_{s}) \ln (R_{c}/R_{s}) }{ 1 + \frac{3}{2} (R_{c}/R_{s}) + 3 (R_{c}/R_{s}) \ln (R_{c}/R_{s}) } \right] E_{\rm int}
    \approx -\frac{8}{3}(1+\chi)^{-1} E_{\rm int},
\ee
as derived in the main text. For a typical range of values of $\chi\approx 0.4-0.6$, this results in a prefactor on the right side of \mbox{$(1+\chi)^{-1}8/3\approx 1.7-1.9$.} Thus the gravitational energy is proportionally smaller than the internal energy for a hot layer in comparison to a self-gravitating body (where this factor would be $2$).



\end{appendix}

\bibliographystyle{aasjournal}

\end{document}